\documentstyle[amssymb,pre,aps,preprint]{revtex}

\begin{document}

\tightenlines

\title{Minority Game: a mean-field-like approach}
\author{In\'{e}s Caridi\thanks{
e-mail caridi@cnea.gov.ar}, Horacio Ceva\thanks{
e-mail ceva@cnea.gov.ar}}
\address{Departamento de F{\'{\i}}sica, Comisi{\'o}n Nacional de
Energ{\'\i }a At{\'o}mica,\\
Avda. del Libertador 8250,1429 Buenos Aires, Argentina }
\maketitle

\begin{abstract}

We calculate the standard deviation of $(N_1 - N_0)$, the difference of
the number of agents choosing between the two alternatives of the minority
game.  Our approach is based on two approximations:  we use the whole set
of possible strategies, rather than only those distributed between the
agents involved in a game; moreover, we assume that a period-two dynamics
discussed by previous authors is appropriate within the range of validity
of our work.  With these appproximations we introduce a set of states of
the system, and are able to replace time averages by ensemble averages
over these states.  Our results show a very good agreement with
simulations results for most part of the informationally efficient phase.

\end{abstract}



\section{Introduction}

The minority game (MG) is an adaptive game introduced by Challet and Zhang
\cite{Challet} to study competitive systems whose available resources are
finite (ecology, financial systems, traffic in Internet, etc).

At every step of the game, the $N$ participants (agents) must choose one of
two alternatives (0 or 1, to buy or to sell, to choose one of two possible
routes, etc.), and the winners will be those who turn out to be in the
minority. Each agent decides her move based on the {\em history} of the game
(a string of $m$ bits with the information about the $m$ previous minority
sides), using one of her set of $s$ prescriptions or {\em strategies}.
Feedback is established by a reward system whereby every winner agent gets a
point; moreover, out of all the strategies in the game, those that correctly
predicted the minority side also get a (so-called virtual) point. The game is
adaptive, because an agent will make her choise using the strategy that at
this particular time has more points.

For every time step, let us call $N_{0}$ ($N_{1})$ the number of agents
choosing side 0 (1), such that $N_{0}+N_{1}=N$. The main variable that is
usually considered is $\sigma $, the standard deviation of the difference ($%
N_{1}-N_{0})$, as a function of $N$, $m$, and $s$,

\begin{equation}
\sigma ^{2}=\frac{1}{T}\sum_{t=1}^{T}(N_{1}-N_{0})^{2}  \label{dispersion}
\end{equation}

where $T$ is the number of steps of the game.

The standard deviation $\sigma $ is a measure of the form in which the
resources (points) of the population are used: whenever $\left|
N_{1}-N_{0}\right| $ is small, more agents get a point, implying a better
utilization of the resources. The model has attracted some attention because
for certain values of $m,$ $s$ and $N$, $\sigma $ turns out to be smaller
than $\sigma _{r},$\ the standard deviation of a random-choise game, where
the $N$ agents choose sides randomly. This result implies that althought
there is no agent-agent interaction in the MG, there is some kind of
emergent coordination; in other words, there is an effective interaction
that appears through global magnitudes of the system. This collective
behavior motivated a diversity of studies of the MG, with a variety of
techniques: numerical simulations \cite{Challet}, \cite{Savit}; mean-field
approximations \cite{Manuca}; equivalence with spin-glass models \cite
{Marsili}; thermal treatments \cite{Cavagna}, etc. Manuca et al. \cite
{Manuca} have studied the complete succession of minority sides, and used a
specific mean-field to calculate $\sigma $. In the present work we calculate
$\sigma $ for the same set of values of $N$, $m$ and $s$ used by these
authors, obtaining a somewhat better agreement with numerical results.

The public information on which strategies are based is the record of the
last $m$ minority sides (that is why $m$ is called the {\em memory} of the
game).  Then, there are ${\cal H=}$ $2^{m}$ different histories, the
history at time $t$ being the string $k=(\chi _{t-m},...,\chi _{t-1})$
where $\chi _{t}$ is the minority side (0 or 1) which results after time
$t$.  In the following we will write $k$ as an integer between 0 and
$2^{m}-1$.  Every strategy is a function that assigns {\em outputs}
(predictions of the following minority side) for every one of the ${\cal
H}$ possible {\em inputs }. Hence, there are ${\cal L}$ ${=2^{{\cal H}}}$
different strategies in a game of memory $m$ ; at the beginning of the
game each agent is randomly assigned $s$ strategies, from the pool of
${\cal L}$ (with replacement).

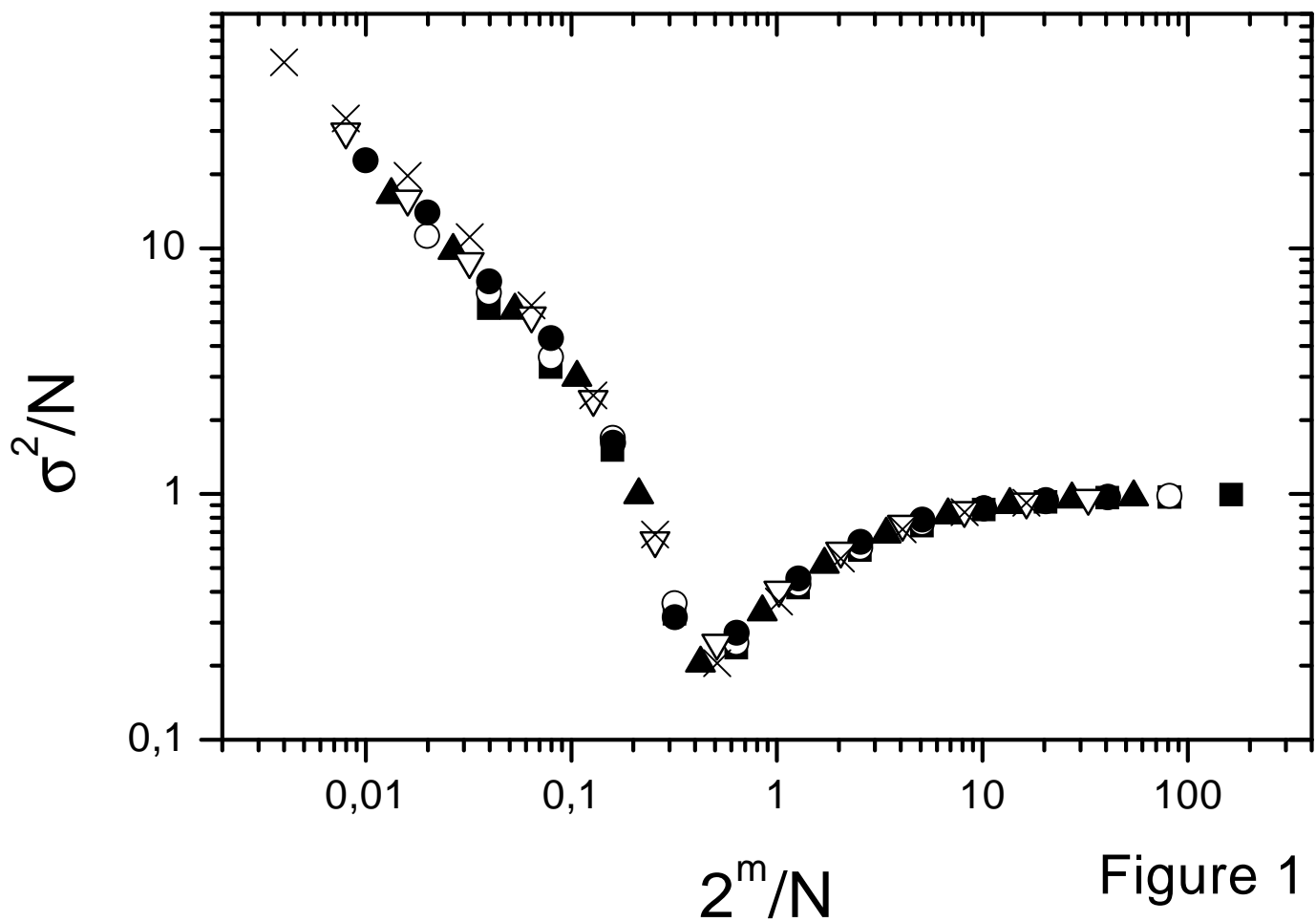
\begin{figure}[tbp]
\caption{Scaling of $\protect\sigma ^{2}/N$ vs $z=2^{m}/N$, for different
values of $m$ and $N$: $\blacksquare $ $N=101$; $\circ $ $N=201$; $%
\blacktriangle $ $N=301$; $\bullet N=401;\bigtriangledown $ $N=501$; $\times
$ $N=1001$. For each value of $N$ and $m$ there are 32 runs, each one of $%
T=10000$ time steps; for $N=1001$, each run has 50000 time steps.}
\label{scaling}
\end{figure}

It is known \cite{Manuca} that data for $\sigma ^{2}/N$ displays scaling for
every value of $s$ when plotted as a function of\ $z=2^{m}/N$. In fact the
plot of $\sigma ^{2}/N$ vs $z$ has two different regions, as it is shown in
Fig. \ref{scaling} : {\em (i)} $z\lesssim 0.5$, which is an informationally
efficient phase. Here $\sigma ^{2}/N$ is a decreasing function of $z,$
showing a change of behavior \ around $\sigma ^{2}/N\approx 1$; {\em (ii)} $%
z>0.5,$ where $\sigma ^{2}/N$ increases and asymptotically approaches the
line $\sigma ^{2}/N=1$. This line corresponds to the random-choise game. Our
calculation is appropriate within a good part of the first region.

Usually, numerical simulations are carried over for a game of T time
steps, and afterwards\ the results are averaged over several independent
runs of the game.  In every run there are ${\ell }\equiv sN$ strategies
distributed between the agents, out of the pool ${\cal L}$.  There are two
points worth noticing:  the pool ${\cal L}$ of {\it all} the strategies is
{\it symmetrical }, in the sense that for every possible history, the
number of strategies predicting $\chi _{t}=1$ is equal to the number of
strategies predicting $ \chi _{t}=0$; on the other hand, every set of
$\ell $ strategies used in each run is only{\it \ approximately}
symmetrical, because it is just a finite sample of ${\cal L}$.  When the
sample is only of a moderate size, however, the difference between the
properties of the sample and the pool is small, and the corresponding
uncertainty is known.  Indeed, this is widely used in standard `sampling'
techniques \cite{Meyer}:  for instance, the uncertainty associated with
samples of size $\thickapprox 400$ is less than $\thickapprox 5\%$.  These
considerations are at the base of our calculation, as we now explain.

In the following we will make analytic calculations over the whole pool $%
{\cal L}$, rather than over samples of size $\ell$. In this form we will
benefit from its symmetry, and also will be able to calculate several
magnitudes\ by simply counting different sets of pairs of strategies (for $%
s=2$). In this sense we are using sampling techniques the other way around:
\ we start from the pool, to find the samples' properties. As it will be
clear below, this is a mean-field-like approach.

Equation (\ref{dispersion}) calculates $\sigma $ as a time average over $T$
time steps. We will replace this by an ensamble average over a (restricted)
set of states of the system. In Sec. II we introduce the states and describe
how a {\it period-two dynamics} observed in Ref. \cite{Manuca} can be used
to substantially reduce the number of states needed. In Sec. III we write
down $\sigma $ in terms of the new variables; to carry out the actual
calculation of $\sigma $ we found it useful to make use of some ad-hoc
diagrams that are also explained. In the final section we offer some
conclusions.

\section{the space of available states}

In this work we will only consider games where $s=2,$ {\it i.e}. every
agent has two strategies to choose from.  To find $\sigma $ from Eq.
(\ref {dispersion}) we need to write down expressions for $N_{1}/N$ and
$N_{0}/N$.  It is clear that the values of these ratios will fluctuate
with the actual numerical realization. We now {\it assume} that we can
obtain approximate values for these variables by using not $N,$ the actual
number of agents, but rather a game with ${\cal N}$, the maximum amount of
(virtual) agents that can originate in the pool. ${\cal N}$ is equal to $%
{{\cal L} \choose 2}%
+{\cal L}$, the number of pairs of strategies that it is possible to make
from the pool ${\cal L}$ (with replacement) \footnote{%
Notice that this excludes the situation where two agents get the same pair.
In any case, the probability that this happens in the game is vanishingly
small even for a moderate value of $m$.} . Within this approximation, we
will calculate $N_{1}/N$ as $\approx {\cal N}_{1}/{\cal N}$ , and $%
N_{0}/N\approx {\cal N}_{0}/{\cal N}$, where ${\cal N}_{0}({\cal N}_{1})$ \
is the number of virtual agents choosing side 0 (1).

We now describe the {\it states }to be used in our ensemble average.
It is convenient to begin by introducing {\em microstates} of the system.
For each time step $t$, we define $a(k^{\prime })\equiv n_{1}(k^{\prime
})-n_{0}(k^{\prime })$, the accumulated difference between the number of
times that after the appearance of string $k^{\prime }$, the resulting
minority side was 1 ($n_{1}(k^{\prime })$) or 0 ($n_{0}(k^{\prime })$). A
microstate $\mu $ is given by $\ (\overrightarrow{a},k_{0})$, where $%
\overrightarrow{a}\equiv (a(0),a(1),...,a({\cal H}-1))$ has the information
of the net amount of virtual points assigned to the pool of strategies, and
$k_{0}$ is the string containing the history that effectively showed up at
time $ t $.

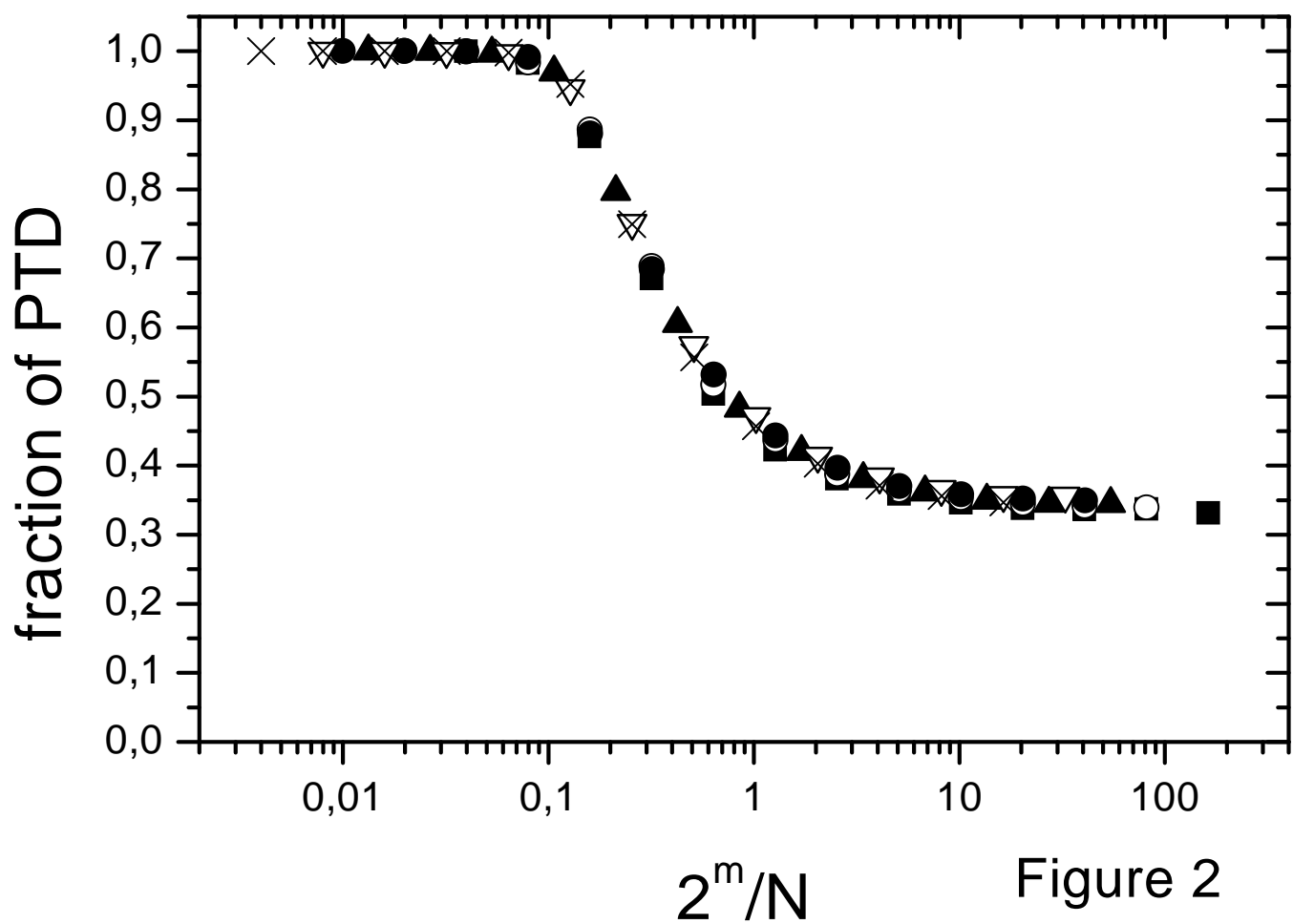
\begin{figure}[tbp]
\caption{Validity of the period-two dynamics. Number of times in which the
dynamics comes true, over the total number of times in which the
corresponding history occurs for an even time. Data are for different values
of $N$: $\blacksquare $ $N=101$; $\circ $ $N=201$; $\blacktriangle $ $%
N=301;\bullet N=401;\bigtriangledown $ $N=501$; $\times $ $N=1001$. For each
value of $N$ and $m$ there are 32 runs, each one of $T=10000$ time steps;
for $N=1001$, each run has 50000 time steps.}
\label{regla_savit}
\end{figure}

The set of microstates is rather complex and very big. In order to get
simpler expressions, we now turn back to the other approximation in which
this work is based on, the period-two dynamics observed in \cite{Manuca}.
Let us consider a game with $s=2,$ $m=2$; in this case there are only four
strings $k:00,$ $01,$ $10$ and $11,$ representing all possible results for
the last two minority groups. In their study of the time series of minority
groups, Manuca et. al. kept a record of all the times when a given string $%
k, $ for instance $k=01$, appeared. This record shows a most remarkable
behavior: odd occurrences of $k$ are followed by a minority side $\chi $\
whose value is essentially random (e.g. $\chi _{odd}=1)$, while for the next
(even) occurrence of $k$ the result is deterministic, being the opposite of $%
\chi _{odd}$ ($\chi _{even}=0$ in this example). This behavior was described
as a period-two dynamics (PTD) by those authors, and can be seen to be
essentially true for $m=2$. In fact, we found that a plot of the probability
that the data follows this rule, as a function of $z=2^{m}/N,$ displays
scaling as can be seen in Fig.\ref{regla_savit}, where one can also see
that this probability is greater than 0.5 if $z\lesssim $ 0.7 .

The simultaneous use of the pool ${\cal L}$ {\it and} PTD allows a dramatic
simplification of the ensemble of states. To understand this point, let us
consider one step $t$ of the game where a minority string $k$ is followed by
a minority side $\chi _{t}$; as mentioned above, ${\cal L}/2$ strategies
predict this output, and therefore get a point. Now, the next time that the
string $k$ appears, because of PTD, the {\it other} half of the strategies
(predicting the opposite minority side) should get a point. Therefore, after
an even appearance of $k$, {\it all} strategies will have one extra point.
But remember that these points are assigned so that one can choose the more
succesfull strategies, simply by picking up that with the greatest amount of
points;\ in this sense, nothing should change if, rather than {\em adding} a
point to the second half of strategies, we{\em \ remove} the point we
already assigned to the first part. The implication of this procedure is
great: we need to consider only those microstates where $\left| a(k)\right|
\leq 1.$ In the following, we will always refer to this subset of
microstates, and the corresponding states.

A state is defined as the set of microstates that have the same value of
the variables\ $m,n,\phi $ and $p,$ where $n=\sum_{\{k^{^{\prime
}}\}}\left| a(k^{^{\prime }})\right| $ , $\phi =\left| a(k_{0})\right| ,$
and $p=a(k_{0})$.  As mentioned above, $\phi $ can take the values 0 or 1,
while $p$ can be equal to 0 or \{1, -1\}, respectively; $\{k^{^{\prime
}}\}$ is the set of all values of $k^{^{\prime }}.$ It is of some help to
think of $(\phi ,p)$ as an spin and its projection.  In the future we will
need to write down the number of microstates of the set, {\em i.e.} the
degeneracy of the state, $ g(n,m,\phi ,p).$

In Fig.\ref{diagrama} we show a diagramatic representation of the
microstates, that is also helpful to visualize the states. Each diagram has $%
{\cal H}$ rows, one for every possible history of $m$ bits; the first column
shows the value of $\overrightarrow{a}$ (the values 1, -1 and 0 are
represented by an up or down arrow, and an empty site, respectively), and
the second column simply displays the actual string $k.$ There can only be
one arrow for row, and up to ${\cal H}$ arrows in the diagram. Hence, the
maximun number of points that the pool ${\cal L}$ can have is $2^{{\cal H}-1}%
{\cal H}$.  \footnote{In the recent past there has been some discussion
about the relevance of the memory \cite{Cavagna-Savit} in the MG. By
randomly generating strings of minority sides, Cavagna obtained results
essentially equal to those of Fig.\ref{scaling}; he concluded that memory
is irrelevant in the MG, and should not be used to explain its behavior.
On the contrary, Savit argued that ``the dynamics and the information
structure in the two versions are fundamentally the same''.  Our diagrams
are particularly useful to shed some ligth on this point.  To choose a
string randomly, simply means to choose in this form the row in the {\it
rigth} column of our diagrams; this eventually will change the
`microscopic' behavior of the model, but the register of what happens
after the appearance of any history is not changed, because the rules that
govern the {\it left} column of the diagram are not changed.}

\section{The standard deviation}

We now have completed the characterization of the states, and can come back
to the calculation of \ (${\cal N}_{1}-{\cal N}_{0})/{\cal N}.$ As our
approach is based on all the pairs of strategies ({\em i.e.} virtual agents)
that can be formed out of the pool , there are three rather natural
magnitudes to consider: ${\cal N}_{d1}$ ( ${\cal N}_{d0}$) the number of
agents whose response to a given microstate will be to choose $\chi =1{\cal %
\ }$($\chi =0$) with certainty, hence the nickname `decided' agents ; and $%
{\cal N}_{u},$ those agents that under the same circumstance can not make up
their minds, thus the nickname `undecided'. Clearly, ${\cal N}$ $=$ ${\cal N}%
_{d1}$ + ${\cal N}_{d0}$ + ${\cal N}_{u}$. In fact, after establishing a
method to know how ${\cal N}_{u}$ will split into those choosing 0 (${\cal N}%
_{u0})$ or 1$\ ({\cal N}_{u1}),$ we can write ${\cal N}_{1}-{\cal N}_{0}=%
{\cal N}_{d1}+{\cal N}_{u1}-{\cal N}_{d0}-{\cal N}_{u0}.$

Recalling that every possible pair of strategies corresponds to one virtual
agent, to obtain the ${\cal N}^{\prime }s$ is convenient to know $E_{1}^{x}(%
\overrightarrow{a},k)$ and $E_{0}^{x}(\overrightarrow{a},k)$, the number of
strategies with $x$ virtual points predicting $\chi =1$ and $\chi =0,$
respectively, as the following minority side after the string $k$. Using $%
E_{1}^{x}$ and $E_{0}^{x}$ as a shortand notation for $E_{1}^{x}(%
\overrightarrow{a},k)$ and \ $E_{0}^{x}(\overrightarrow{a},k)$, it is

\begin{eqnarray}
{\cal N}_{u} &=&\sum\limits_{x}E_{1}^{x}E_{0}^{x}  \nonumber \\
{\cal N}_{d1} &=&\sum\limits_{x}\sum\limits_{j<x}E_{1}^{x}E_{0}^{j}+\frac{1}{%
2}\sum\limits_{x}\sum\limits_{j\neq x}E_{1}^{x}E_{1}^{j}+\sum\limits_{x}{%
(E_{1}^{x})}^{2}  \nonumber \\
{\cal N}_{d0} &=&\sum\limits_{x}\sum\limits_{j<x}E_{1}^{j}E_{0}^{x}+\frac{1}{%
2}\sum\limits_{x}\sum\limits_{j\neq x}E_{0}^{x}E_{0}^{j}+\sum\limits_{x}{%
(E_{0}^{x})}^{2}  \label{enes}
\end{eqnarray}

The last two terms of ${\cal N}_{d1}$ and ${\cal N}_{d0}$ are equal to ${{%
C_{2}^{{\cal L}/2}+}}{\cal L}/2$ (where $C_{w}^{q}\equiv
{w \choose q}%
$ $)$, and are related with those agents for which both of their strategies
predict the same side for the string $k$. As we are interested in the
difference (${\cal N}_{1}-{\cal N}_{0}),$ these factors will cancel out, and
don't need to be considered in the following.

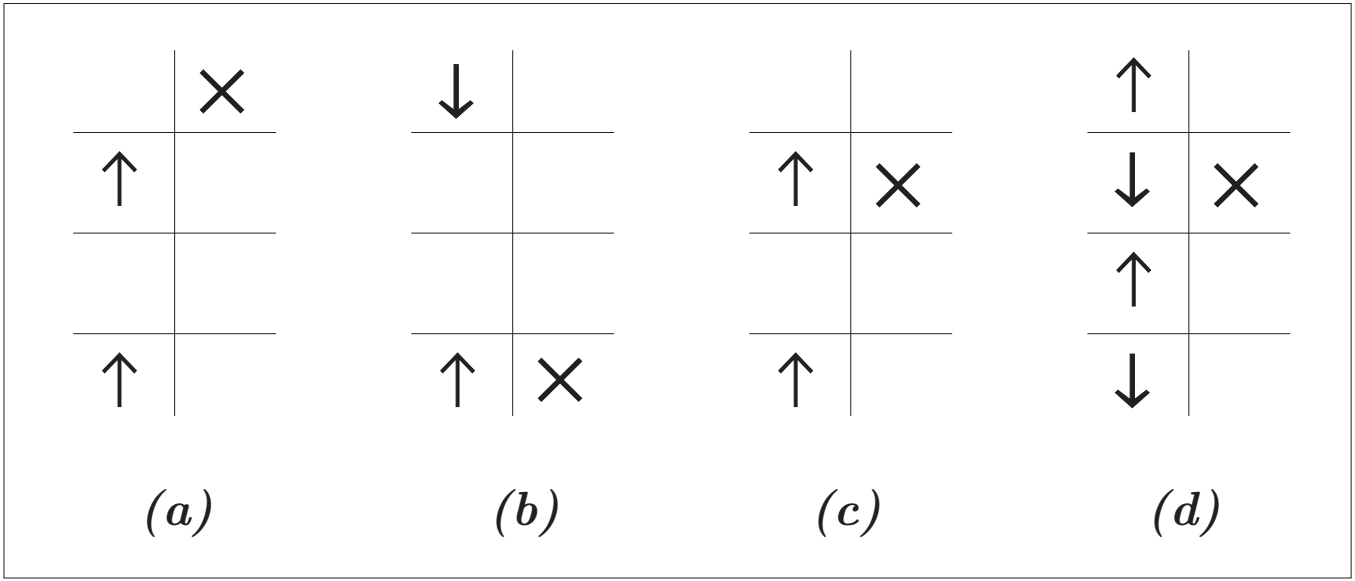
\begin{figure}[tbp]
\caption{Diagrams used to represent the ensamble of microstates for $m=2$.
>From top to bottom, the rows correspond to the strings 00, 01, 10 and 11.
The symbol $ \times $ is used to indicate the actual history; the arrows
are used to represent the virtual points assigned to the strategies.  For
every history with an arrow, there are ${\cal L}/2$ points assigned.  $a$:
$(2,2,0,0)$; $b$ :  $(2,2,1,1)$; $c$:  $(2,2,1,1)$; $d$:  $(2,4,1,-1)$.
Notice that $b$ and $c$ are two microstates corresponding to the same
state.}
\label{diagrama}
\end{figure}

Microstates $\mu _{1}=(\overrightarrow{a}_{1},k_{1})$, $\ \mu _{2}=(%
\overrightarrow{a}_{2},k_{2})$ obtained one from the other by interchanging
one or more rows in the diagrams of Fig.(\ref{diagrama}) \ are symmetrical,
in the sense that they have the same values $E_{i}^{x}$: $E_{i}^{x}(\mu
_{2})=E_{i}^{x}(\mu _{1})$ , $(i=0$ or $1)$.This, in turn, implies ${\cal N}%
_{u}(\mu _{2})={\cal N}_{u}(\mu _{1}),$ and similarly for all terms of Eq.(
\ref{enes}). In other words, the values of these magnitudes depend only on $%
(m,n,\phi ,p),$ i.e are state dependent. Therefore, we will only need to
consider the $E^{\prime }s$ for different {\em states, }\ together with the
corresponding degeneracies $g(m,n,\phi ,p).$ Thus, the expression for the
standard deviation becomes

\begin{equation}
\sigma ^{2}=\frac{N^{2}}{{\cal N}^{2}}\frac{1}{\Omega }\sum_{\{m,n,\phi
,p\}}g(m,n,\phi ,p)\left[ {\cal N}_{d1}+{\cal N}_{u1}-{\cal N}_{d0}-{\cal N}%
_{u0}\right] ^{2}  \label{dispersion2}
\end{equation}

where $\Omega =\sum_{\{m,n,\phi ,p\}}g(m,n,\phi ,p).$

For every state, we assume that each undecided agent chooses randomly
between $\chi =0$ and $\chi =1.$ Moreover, $\sigma ^{2}$ will be averaged
over a certain number of independent runs; hence, we only need the average
values of ${\cal N}_{u1}$ and ${\cal N}_{u1}^{2},$ that are given by the
standard expressions of the random walk, $\langle {\cal N}_{u1}\rangle =%
\frac{1}{2}{\cal N}_{u}$ and $\langle {\cal N}_{u1}^{2}\rangle =\frac{1}{4}%
{\cal N}_{u}+\frac{1}{4}{\cal N}_{u}^{2}$.

Then

\begin{equation}
\sigma ^{2}=\frac{N^{2}}{{\cal N}^{2}}\frac{1}{\Omega }\sum_{\{m,n,\phi
,p\}}g(m,n,\phi ,p)\left[ {\cal N}_{u}+({\cal N}_{d1}-{\cal N}_{d0})^{2}%
\right]  \label{dispersion3}
\end{equation}

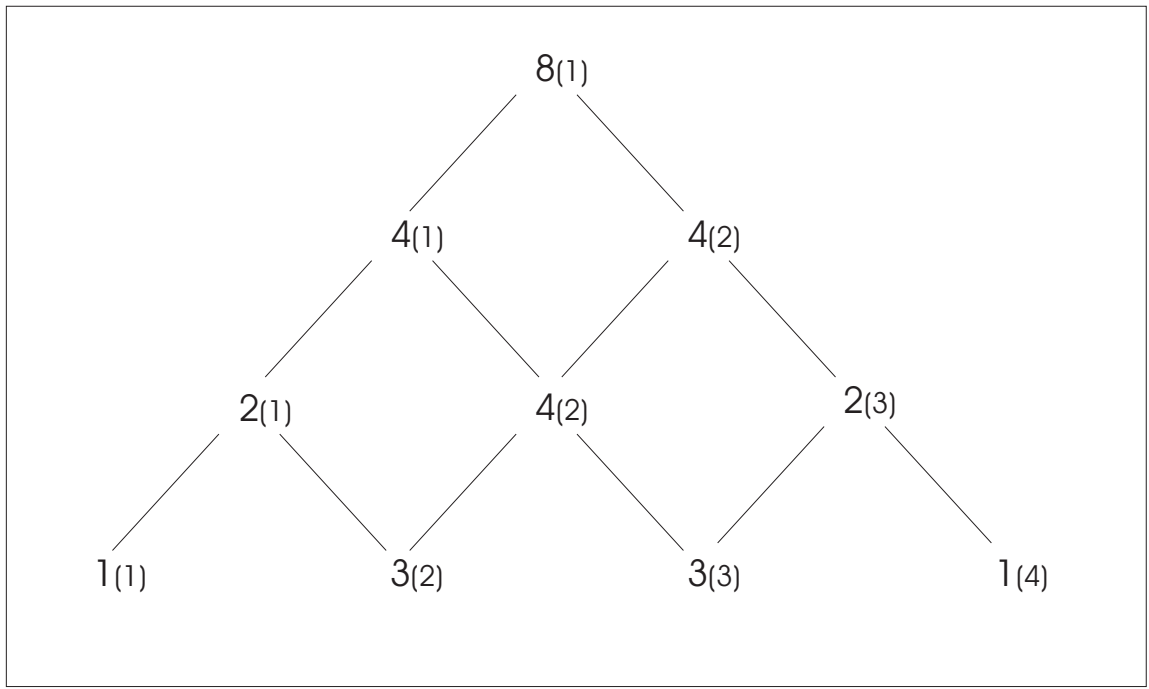
\begin{figure}[tbp]
\caption{Array of strategies corresponding to states with $m=2$, $\protect
\phi=1$, and $p=-1$.  In this tree there are represented the ${\cal
L}/2=8$ strategies predicting $\chi =0$ for the actual string.  Each row
of the tree is characterized by a value of $n$, beginning with $n=1$, up
to $ n=4$.  In each site of the tree the number between brackets gives
the amount of virtual points ($x=1,..,n$) for the corresponding
strategies.  There is an analogous tree for the strategies predicting
$\protect\chi =1$ for the same string.  The difference of both trees is
given only for the virtual points assigned; in the tree for
$\protect\chi=1$ these values are displaced, so that each row begins with
$x=0$, and ends up with $x=n-1$.}
\label{Tartaglia}
\end{figure}

The calculation of the different terms of Eq.(\ref{enes}), and the
corresponding calculation of $\sigma $ (Eq.(\ref{dispersion3})) is
straigthforward, but rather lengthy and cumbersome.  We have displaced to
an Appendix some of the details;\ on the other hand it is useful to use a
graphic representation to find $E_{0}^{x},$ $E_{1}^{x}$ , and show how to
obtain these terms.  Notice that, regardless of which microstate we are
looking at, the set of $all$ possible strategies will always split into
two groups of ${\cal L}/2$ strategies each, predicting $\chi =0$ or $1$
respectively; therefore, we just have to find how each one of these groups
will split into smaller subgroups characterized by the number of points.
Let us consider, to be specific, the case $m=2$.  In this case, we only
need to know the $E$'s for ($n=0$ to $3,$ $\phi =0,$ $p=0),$ and for
($n=1$ to $4,$ $\phi =1,$ $p=\pm 1).$ In Fig.(\ref{Tartaglia}) we show \
these values for the case $\phi =1,$ $p=-1,$ for strategies predicting
$\chi =0$ as the following minority side.  Being $\phi =1$ implies that we
have to consider microstates having $a(k_{0})\neq 0$.  The simplest form
to find the $E$'s in this case, is by using the diagrams with one to four
arrows.  Thus, it is clear that there are ${\cal L}/2$ strategies with one
point ($and$ predicting $\chi =0$).  With $\left|
\overrightarrow{a}\right| =2$ (i.e.  $n=2$), there will be ${\cal L}/4$
strategies predicting only one of the minority sides (having one point),
and ${\cal L}/4$ predicting both minorities correctly (thus having two
points); analogously, for $n=3$ there will be three groups of ${\cal
L}/8$, ${\cal L}/4,$ and ${\cal L}/8$ strategies, with 1, 2 and 3 points,
respectively; finally, for $n=4$ there are four groups with ${\cal L}
/16,$ ${\cal L}(1+2)/16,$ $\ {\cal L}(1+2)/16$ and ${\cal L}/16$, \ with
1, 2, 3 and 4 points respectively.  Graphically, this gives rise to a
tree, as that shown in Fig.\ref{Tartaglia}, in which every row corresponds
to a given value of $n,$ and in every site of the tree one writes
$E_{1}^{x}.$ In fact, introducing $e_{i}^{x\text{ }}$ by writing
$E_{i}^{x}(m,n,\phi ,p)=2^{{\cal H }-n+\phi -1}e_{i}^{x}(m,n,\phi ,p),$
with $i=1$ or $0,$ it is possible to see that the tree associated with
$e_{i}^{x}$ turns out to be Tartaglia's triangle, whose analytic
properties are well known.  The cases including states where $(n=\phi =0)$
\ can be handled in the same form.  The difference between states with $\
\phi =0$ and $\phi =1$ it is apparent in the schemes (a) and (c) of
Fig.\ref{diagrama}, but from the point of view of the trees, it simply
implies that when $\phi =0$ the values of $n$ are `shifted'; thus the \
`upper' row has $n=0$, and the \ `lowest' one has $n={\cal H}-1;$ moreover
in row $n$, say, the different terms will have 0,1,...,$n$ points.  All
terms needed to calculate $\sigma $ are simply related with different sums
and products of terms of these trees, without mixing factors from
different rows.

To carry out the calculation of $\sigma ^{2}$ we use \ ${\cal N}_{u}$ , $%
\left| {\cal N}_{d1}-{\cal N}_{d0}\right| {\cal \ }$\ and $g$, as obtained
in the Appendix. The actual expressions for ${\cal N}_{u}$ and $\left| {\cal %
N}_{d1}-{\cal N}_{d0}\right| $ are functions of the value of $\phi ,$ namely

\begin{eqnarray}
{\cal N}_{u} &=&\left( 2^{{\cal H}-n-1}\right) ^{2}\text{ }C_{2n}^{n}
\nonumber \\
{\cal N}_{d1}-{\cal N}_{d0} &=&0  \label{final1}
\end{eqnarray}

for $\phi =0,$ and

\begin{eqnarray}
{\cal N}_{u} &=&C_{2n-2}^{n}\text{ \ \ \ \ if \ \ }n>1  \label{final1b} \\
{\cal N}_{u} &=&0\text{ \ \ \ \ \ \ \ \ \ \ \ if \ \ }n=1  \nonumber \\
\left| {\cal N}_{d1}-{\cal N}_{d0}\right| &=&C_{2n-2}^{n}+C_{2n-2}^{n}
\nonumber
\end{eqnarray}

\bigskip for $\phi =1$.

The state degeneracy is \footnote{%
in fact this expression is totally correct in the case where the microstates
have a uniform probability distribution; in a game with memory the
succession of strings is obviously concatenated and Eq.(\ref{degeneration})
is only approximated}

\begin{equation}
g(m,n,\phi ,p)=C_{{\cal H}}^{n-\phi }({\cal H}-n+\phi )\text{ }2^{n-\phi }
\label{degeneration}
\end{equation}

where $\ \phi \leq n\leq {\cal H+}\phi -1$

Two points are worth mentioning, in relation with $\sigma ^{2}.$ First,
using Eq.(\ref{N1-N0}) in Eq.(\ref{dispersion3}), we verify that $\sigma ^{2}
$ can be written just in terms of ${\cal N}_{u}$ 's factors. Furthermore,
this expression can be factored out as follows: $\sigma ^{2}=N$ ${\cal F}
\left( m\right) $. Notice that our general assumptions require to consider
$N$ big enough, so that the replacement of the samples by the whole pool
makes sense; hence, the factorization should also be valid in the same
limit.

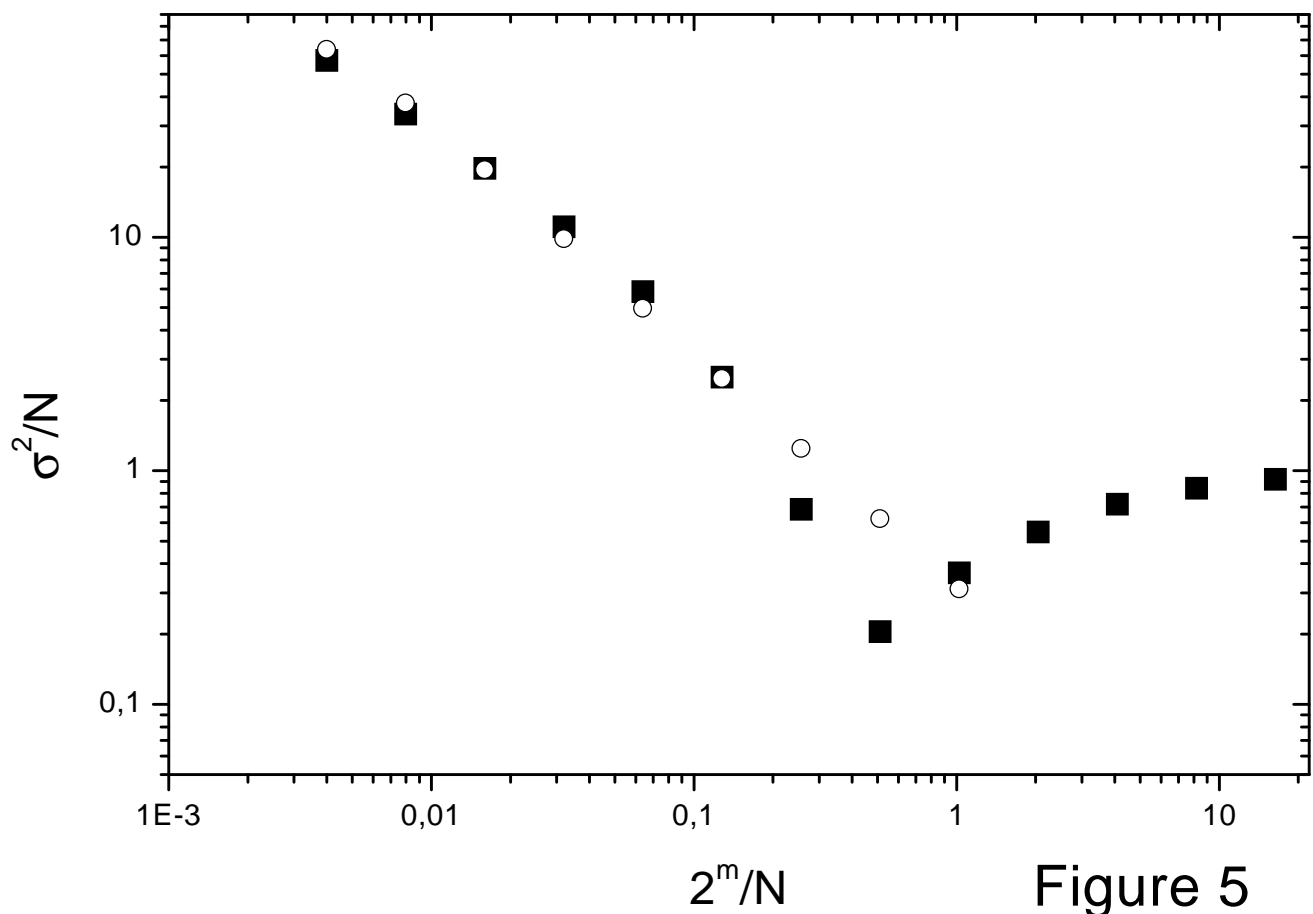
\begin{figure}[tbp]
\caption{Data of $\protect\sigma ^{2}/N$ vs $z=2^{m}/N$ for $N=1001$, both
for the numerical simulations and from our calculations. As before, the
results are averaged over 32 runs of $50000$ time steps each.}
\label{calculo}
\end{figure}

Our results are presented in Fig.\ref{calculo}. We have done numerical
simulations of the MG for $N$ from $101$ up to 1001, 50000 time steps, and $%
m=2$ up to 13; the results where averaged over 32 independent runs. We show
both the usual simulation data, as well as the results of the calculation
with the equations derived in this work. As it can be seen, if $\frac{{%
\sigma }^{2}}{N}\geq 1$ (and $2^{m}/N$ $\leq 0.1)$ there is an excellent
agreement between these two sets of data. For $2^{m}/N$ between 0.1 and 0.7
we only have a qualitative agreement, and for $2^{m}/N>0.7$ our results (not
shown) are clearly inadequate.This coincides with the range of validity of
the results of Manuca et al. \cite{Manuca}, i.e we only cover the
informationally efficient phase. It is to be noticed that in the first
region our results can be fitted to a high degree of accuracy by a straight
line with slope -1.

\section{conclusions}

We have made a mean-field-like calculation of $\sigma $, the standard
deviation of the MG. Our approach is based on two approximations:  on the
one hand, rather than using the properties of the actual samples used to
calculate, we consider the whole pool of strategies, ${\cal L}$; in this
form we benefit from the symmetry of the histories.  It should be noticed
that we actually quote results that are averages over several independent
runs, so that it is reasonable to expect that they should be near to the
symmetry of the pool.  On the other hand, we considered that the
period-two dynamics introduced by Manuca et.al \cite{Manuca} is correct
within the range of validity of our work.  We used these approximations to
replace the time averages appearing in Eq.(\ref{dispersion}), by ensemble
averages over the states.

Our results show an excellent agreement with data from the simulation in the
region  $\sigma ^{2}/N\gtrsim 1.$ The calculation embraces an important part
of the informationally efficient region, but the agreement is lost when $%
\sigma ^{2}/N$ is near of its minimun value.

This method can also be useful to deal with cases where $s=3$ (one needs to
consider all the sets of three strategies) or greater.

H.C. was partially supported by EC grant ARG/b7-3011/94/27, Contract 931005
AR.

\appendix

\section*{}

We provide some extra details about the calculation of ${\cal N}_{u}$ , $%
\left| {\cal N}_{d1}- {\cal N}_{d0}\right|$ and $g$.

The values of $e_{i}^{x},$ (for $i=0,1$), that we have already identified
with the coeficients in Tartaglia's triangle, are state dependant (remember
that in site $(q,w)$, the coeficient of the tree is the combinatorial $%
C_{w}^{q}\equiv
{w \choose q}%
.$

\begin{mathletters}
\label{allequations}
\begin{eqnarray}
e_{0}^{x}(m,n,0,0) = e_{1}^{x}(m,n,0,0)=C_{n}^{x}  \label{equationa}
\end{eqnarray}
\begin{eqnarray}
e_{0}^{x}(m,n,1,1) = e_{1}^{x}(m,n,1,-1)=C_{n-1}^{x}  \label{equationb}
\end{eqnarray}
\begin{eqnarray}
e_{0}^{x}(m,n,1,-1) = e_{1}^{x}(m,n,1,1)=C_{n-1}^{x-1}  \label{equationc}
\end{eqnarray}

where $0\leq x\leq n$, $0\leq x\leq n-1$, and $1\leq x\leq n$ in Eq.(\ref
{allequations}a-c), respectively. In the first equation $\phi =0$ and
therefore $0\leq n\leq {\cal H}-1$, while the last two equations correspond
to states with $\phi =1$, then $1\leq n\leq {\cal H}$.

We begin with $\phi =0$ states, i.e. $(m,n,0,0)$ states, for which,
following Eq.(\ref{enes})

\end{mathletters}
\[
{\cal N}_{u}=2^{2({\cal H}-n-1)}\sum_{x=0}^{n}e_{1}^{x}e_{0}^{x}=%
\sum_{x=0}^{n}(C_{n}^{x})^{2}=C_{2n}^{n}
\]

Moreover, for the same states

\begin{eqnarray}
{\cal N}_{d1}=2^{2({\cal H}-n-1)}\sum_{x=0}^{n}%
\sum_{j=0}^{n}e_{1}^{x}e_{0}^{j}=2^{2({\cal H}-n-1)}\sum_{x=0}^{n}%
\sum_{j=0}^{n}C_{n}^{x}C_{n}^{j}={\cal N}_{d0}  \nonumber
\end{eqnarray}

Therefore, $\left| {\cal N}_{d1}-{\cal N}_{d0}\right| =0.$

Let us now consider $\phi =1,p=1$ states, i.e. $(m,n,1,1)$ states

\begin{eqnarray}
{\cal N}_{u}=2^{2({\cal H}-n)}\sum_{x=1}^{n-1}e_{1}^{x}e_{0}^{x}=2^{2({\cal H%
}-n)}\sum_{x=1}^{n-1}C_{n-1}^{x}C_{n-1}^{x-1}=C_{2n-2}^{n}  \nonumber
\end{eqnarray}

for $n>1$. If $n=1$, it is ${\cal N}_{u}=0$.

${\cal N}_{d1}$ and ${\cal N}_{d0}$ are given by

\[
{\cal N}_{d1}=2^{2({\cal H}-n)}\sum\limits_{j<x}e_{1}^{x}e_{0}^{j}=2^{2(%
{\cal H}-n)}\sum_{j=0}^{n-1}\sum_{x=j+1}^{n}C_{n-1}^{j}C_{n-1}^{x-1}
\]

\begin{eqnarray}
{\cal N}_{d0}=2^{2({\cal H}-n)}\sum\limits_{x<j}e_{1}^{x}e_{0}^{j}=2^{2(%
{\cal H}-n)}\sum_{x=1}^{n-2}\sum_{j=x+1}^{n-1}C_{n-1}^{x-1}C_{n-1}^{j}
\nonumber
\end{eqnarray}

Finally,

\begin{eqnarray}
\left| {\cal N}_{d1}-{\cal N}_{d0}\right|
=\sum_{j=0}^{n-1}C_{n-1}^{j}C_{n-1}^{j}+%
\sum_{j=0}^{n-1}C_{n-1}^{j}C_{n-1}^{j+1}=C_{2(n-1)}^{n-1}+C_{2n-2}^{n}
\nonumber
\end{eqnarray}

More explicitely,

\begin{eqnarray}
\left| {\cal N}_{d1}(m,n,1,p)-{\cal N}_{d0}(m,n,1,p)\right| = {\cal N}%
_{u}(m,n-1,0,0)+{\cal N}_{u}(m,n,1,p)  \label{N1-N0}
\end{eqnarray}

It is remarkable that the difference $\left| {\cal N}_{d1}-{\cal N}%
_{d0}\right| $ can be written just in terms of ${\cal N}_{u}$.

Notice that this `nasty' collection of coefficients can be identified with
combinations of elements of the trees. The simplest case is ${\cal N}_{u}$
in Eq.(\ref{enes}); in this case both trees (for $\chi =0$ or $1$) are
equal, and ${\cal N}_{u}$ is given by the sum of the squares of the factors
in row $n$ of the tree.

To calculate the degeneracies $g(m,n,\phi ,p)$, one has to count the number
of microstates sharing the same values of $(m,n,\phi ,p)$, using the
diagramatic representation of the states. For $\phi =0$,

\begin{equation}
g(m,n,0,0)=C_{{\cal H}}^{n}\text{ }({\cal H}-n)\text{ }2^{n}
\end{equation}

where $\ 0\leq n\leq {\cal H}-1.$ The first factor counts the number of ways
that $n$ arrows can be distributed into ${\cal H}$ places, the second is the
number of strings available for being the actual history ($\times $ in the
diagrams), and the third takes into account that each arrow can be up or
down.

Analogously, we can write $g$ for the case $\phi =1$

\begin{equation}
g(m,n,1,\pm 1)=C_{{\cal H}}^{n}\text{ }n\text{ }2^{n-1}
\end{equation}

with $\ 1\leq n\leq {\cal H}$.

It is possible to write these expressions in a form valid for both values of
$\phi $, as in Eq.(\ref{degeneration}).

Finally, the sum over all the degeneracies is the total amount of
microstates, $\Omega (m)=\sum_{\{m,n,\phi ,p\}}g(m,n,\phi ,p)=3^{2^{m}}2^{m}$%
.

\end{document}